\documentclass{article}
\usepackage{biblatex}  
\usepackage{amsmath}
\usepackage{arxiv}

\usepackage[utf8]{inputenc} % allow utf-8 input
\usepackage[T1]{fontenc}    % use 8-bit T1 fonts
\usepackage{hyperref}       % hyperlinks
\usepackage{url}            % simple URL typesetting
\usepackage{booktabs}       % professional-quality tables
\usepackage{amsfonts}       % blackboard math symbols
\usepackage{nicefrac}       % compact symbols for 1/2, etc.
\usepackage{microtype}      % microtypography
\usepackage{lipsum}
\usepackage{graphicx}
\graphicspath{ {./images/} }

\title{Parallel and real-time post-processing for quantum random number generators}

\author{
 Xiaomin Guo,Fading Lin,Jiehong Lin,Zhijie Song,Yue luo,Qiqi Wang,Yanqiang Guo$^{*}$\\
$^{*}$ Corresponding author:guoyanqiang@tyut.edu.cn
}

\begin{document}
\maketitle
\begin{abstract}
Quantum systems are particularly suited for generating true randomness due to their inherent unpredictability, which can be justified on physical principles. However, practical implementations of Quantum RNGs (QRNGs) are always subject to noise, or uncontrollable influences, diminishing the quality of raw randomness produced. This necessitates post-processing to convert raw output into genuine randomness. In current QRNG implementations, the critical issue of seed updating is often overlooked, risking security vulnerabilities due to increased security parameters when seeds are reused in post-processing, and frequent seed updates fail to yield net randomness, while reusing seeds relies on the assumption that the original sequence inputs are independent.In this work, we have provided a specific scheme for seed updates that balances practicality and security, exploring the parallel and real-time implementation of multiple seed real-time updating toeplitz hash extractors in an FPGA to achieve parallel QRNGs, focusing on efficient hardware computation resource use. Through logic optimization, we achieved a greater number of parallel channels and a post-processing matrix size three times larger than previous works on the same FPGA platform, utilizing fewer logic resources. This resulted in a higher rate of random number generation and enhanced security. Furthermore, with the use of higher-performance ADCs, we attained a random number production rate exceeding 20Gbps.High-speed random number transfer and seed updating were achieved using the PCIe high-speed interface.This marks a significant step toward chip-based parallel QRNGs, enhancing the practicality of CV QRNGs in trusted, device-independent, and semi-device-independent scenarios.
\end{abstract}

% keywords can be removed
%\keywords{First keyword \and Second keyword \and More}

\section{Introduction}
Random number is an essential resource in many modern applications such as cryptography [1,2], statistics, scientific simulation[3], lottery, etc.Quantum random number generators (QRNGs) exploit inherent uncertainty essence of quantum sources and are able to generate unpredictable, non-reproducible true random numbers, even to an agent who holds some side information. Randomness extraction or post-processing is an indispensable strategy for preventing opponents’ attacks by using classical or quantum side information. There have been many ways of random number post-processing involving XOR operation [5,6], Von Neumann’s post-processing[7], linear feedback shift register [8], Trevisan extractor[4,9,10], toeplitz-Hash extractor[4,11-15], etc.Among them, post-processing based on the Trevisan extractor and Toeplitz hash extractor are two strong extractors and information theoretically provable,which have long been applied in privacy amplification and transfer into QRNG to solve the similar questions[9,15]. Between the two, Toeplitz hash extractor has a significantly faster speed thanks to its relatively simple structure[4].
  
   In the early works, post-processing based on Toeplitz matrix operation was implemented on the platform of computer. The rates of production previously reported attained 68Gbps[12] and 100Gbps[22]. In those works, an ADC or oscilloscope is used to collect and save data in a computer hard disk, and then implement the post-processing algorithm using a software like Visual Studio or MATLAB. In this scheme, the speed of random number generation is normally defined as the speed at which bits are generated using software, in other word, it is equivalent off-line rate, and the post-processing is essentially off-line post-processing. The other type is real-time post-processing based on field-programmable gate array (FPGA) reported recent years[13-15]. Compared to offline post-processing based on computer, FPGA-based randomness extractor extracts and outputs true random numbers while ADC acquiring raw random numbers, which is desiderated in practical cryptographic applications for its immediate security and integration. Real-time productive rate up to 6 Gbps[13] ,8.25Gbps[14] and 18.8Gbps[15] have been reported recently. It must be much lower than the equivalent off-line rate mentioned above due to limited hardware source such as clock, logical resource, et.al. The performance of their productive rate inevitably rests with the hardware resource such as ADC acquisition rate or FPGA logical resource, which naturally corresponds to high economic cost[13,15]. Oriented towards practical application, especially when QRNG are gradually accessing in market application like mobile phones, bank USB sticks, driverless cars and other Internet of Things products, overall arrangement and well exploitation of hardware sources have to be fully considered in the construction of random number extractor.
   
Randomness extractors are designed to distill uniform randomness from entropy sources based on random seeds[9]. In the majority of implementations of QRNGs to date, the issue of seed updating has not been addressed, a problem that can no longer be ignored because reusing the same seed in post-processing could lead to an increase in the security parameter[20], thereby introducing potential security vulnerabilities. On one hand, the length of the seed used to construct Toeplitz hash matrices exceeds the length of the true random number sequence generated after one operation. Consequently, updating the seed with each operation fails to extract net randomness[4]. So, updating seed for each time operation is a Pyrrhic. Another existing suggestion posits that the raw data inputted in each instance should be mutually independent, under which premise the same seed may be reused. However, it is imprudent to make such an assumption in this context, as limited detection bandwidth inevitably introduces temporal correlations between samples[21]. On the other hand, the issue of collision probability in Toeplitz hashing arises when two distinct input strings are hashed to the same output string, which is an inherent possibility in any hash function due to the pigeonhole principle. In the context of QRNG and specifically within Toeplitz post-processing, the collision probability is critical because it can impact the unpredictability of the generated random numbers. Under the condition of the same Toeplitz hash extraction ratio, the larger the scale of the matrix, the smaller the collision probability and the higher the security, which requires us to implement a sufficiently large-scale Toeplitz matrix in the FPGA as much as possible[13]. When confronted with these realistic issues, real-time, high rate, seed renewal and collision probability, more comprehensive solution need to be devised in the hardware implementation of the Toeplitz hash post-processing in QRNG.

\section{Security and the allocation of hardware source}
\subsection{\textbf{}\textbf{ Basis from information theory}
}

A randomness extractor is essentially a function. An extractor constructed from a d-bit seed outputs an almost perfect m bits random string by multiplying an n bits partial random raw data. In privacy amplification, the seed source is assumed to be free[4]. whereas in the extractor, seed consumption must be considered, the creed in such field is to use as small a seed as possible to generate as long a random number as possible[23]. As a strong extractor, Toeplitz extractor retain the ability to output random sequence with near uniform distribution when reuse a seed, but cannot ignore a consistently gaining security parameter, especially for a long running RNG. 

In cryptography, the deviation between practical and ideal protocols can be characterized by a security parameter  $\varepsilon$, and statistical distance serves as a standard security measurement. The formula is as follows:

\begin{equation}
\left\| {Z - Y} \right\| \equiv \mathop {\max }\limits_{X \subseteq T} \left| {\sum\limits_{x \in X} {(prob\left[ {Z = x} \right] - prob} \left[ {Y = x} \right]} \right| \buildrel\textstyle.\over= \frac{1}{2}\sum\limits_{x \in X} {(prob\left[ {Z = x} \right] - prob} \left[ {Y = x} \right] \le \varepsilon 
\end{equation}

For two probability distributions X and Y defined on the same domain T, if their statistical distance is bounded by $\varepsilon$, X and Y are said to be $\varepsilon$-close,  that is to say, apart from a small probability $\varepsilon$, it is practically impossible to distinguish between X and Y. The statistical distance quantifies the distinguishability between two probability distributions. In the context of designing QRNGs, the output of an actual QRNG is referred to as $\varepsilon$-close to ideal true random numbers. In actual operation, the composability of security parameters should be considered, in which security parameters for hash functions and the seeds are factors[20]:

\begin{equation}
\varepsilon = N\times \varepsilon_{hash}+\varepsilon_{seed}
\end{equation}

N represents the number of post-processing operations performed using the same seed.
Suppose a bit string Z contains the number of elements represented by \emph{z}; After the action of the hash function, a bit string V is obtained, and its number is represented by \emph{v}. The set of hash functions is represented by G, and g is used to represent one of the hash functions, g$\in$G . When two distinct elements,  and , in set Z become equal after being processed by a hash function, i.e. g($z_a$ )=g($z_b$), it is referred to as a collision:

\begin{equation}
[p\{ g({z_a}) = g({z_b})|{z_a} \ne {z_b}\}  \le \frac{\mu }{{|V|}}
\end{equation}
In the formula, $\mu $ represents universality. When $\mu $ equals 1, the hash function is a universal hash function. In this paper, one of the two universal hash functions used in the post-processing of the quantum random number generator is the toeplitz hash function. A certain toeplitz hash collision probability is equal to $m \times {2^{( - n + 1)}}$[13],it can be seen that under the condition of the same toeplitz hash extraction ratio, the larger the scale of the matrix, the smaller the collision probability and the higher the security, which requires us to implement a sufficiently large-scale toeplitz matrix in the FPGA as much as possible. 

Extraction ratio of true random numbers from raw data is realized by the relative length of row and column of the toeplitz matrix, and theoretically depends on the minimum entropy and security parameter $\varepsilon_{hash}$. Furthermore, the upper limit of matrix dimension can be constructed in FPGA is determined by hardware resource, the first is the logic resources. According to the information theory, extraction ratio of true random number in post-processing is constrained by the Leftover Hash Lemma [16]:
\begin{equation}
    m \le n \times {H_{\min }} - \log \frac{1}{{{\varepsilon _{hash}}^2}}
\end{equation}

where $H_{min}$ is the quantum conditioned minimum entropy, the maximal amount of extractable randomness in presence of classical and quantum side information[17]. $\varepsilon_{hash}$ is the hash security parameter which refers to the proximity of the distribution of extracted random numbers to the uniform distribution.

In quantum random number generation (QRNG), a lower security parameter enhances security but reduces the rate of random number production, necessitating a balance between security and efficiency. Table 1 presents the security parameters chosen for various QRNG projects as reported in the literature. Security parameters increase with more post-processing iterations, suggesting a need for a threshold-based approach for updating the random seed based on post-processing frequency. At the onset of operations, a reduced security parameter can be assigned for post-processing, and upon this parameter diminishing to a pre-set threshold, the seed is renewed via PCIe. This strategy ensures the continuity of QRNG operations without hindering seed updates.  In our research, the QRNG system refreshes its seed every 24 hours, starting with a security parameter of ${10^{ - 50}}$  and updating when it reaches ${10^{ - 36}}$  . The seeds are derived from a prior QRNG design with a security parameter of ${10^{ - 50}}$. This method maintains a balance between continuous operation and security, reflecting both practical deployment considerations and quantum randomness theoretical frameworks.

In our work, the original random sequence of four sidebands acquired by the 16-bit ADC is computed with quantum conditional minimum entropy of 13.0, 13.13, and 13.21With choosing a security parameter of the matrix sizes of the four channels are set 1729×2464, 1729×2464 ,1729×2432 and1729×2432.

\begin{table}[ht]
\caption{Comparison of selected security parameter in QRNGs}
\begin{center}
\setlength\arrayrulewidth{1pt}
\begin{tabular}{c c  } 
 \hline
 Ref. year & Security Parameter \\ [0.5ex] 
 \hline
 [14],2019 & $2^{-50}$ \\

[19],2019 & 3.8×$10^{-10}$\\

 [20],2021 & $10^{-32}$, $10^{-9}$ after 10 years \\

[24],2020 & $2^{-50}$\\

 This work &$10^{-49}$, $10^{-36}$ as threshold \\\hline
\end{tabular}
\end{center}
\vspace{-1.5em}%
\end{table}
\subsection{\textbf{}\textbf{ RESOURCE ALLOCATION}
}

According to the theoretical analysis of post-processing security in information theory [25], the m and n of toeplitz matrices should be at least $10^{2}$ in order of magnitude, and the larger the post-processing matrix scale, the safer the extraction. In our work, we make efforts to coordinate the logic resources inside a FPGA, aiming to achieve as many channel routes as possible and scale the matrix to the fullest extent. Based on progressive elaboration with regards to security , parallel computation of four toeplitz matrices with dimensions approximately 1729*2464 on Xilinx Kintex7-325t FPGA can fully leverage logic resources. Under these operating conditions, we used fewer resources to build a 4-channel post-processing block, and the toeplitz matrix size was 3 times larger than in previous work. Given the resource constraints of FPGAs, it is impractical to directly implement such a large complete matrix extractor on the FPGA without modification. In addition, due to the limitation of ADC bandwidth, it is also necessary to generate n-bit original random bits for a certain clock period, which is in line with our concurrent flow operation idea, in order to reduce the use of resources, there is no need to completely wait for the generation of n-bit original bits. We decompose the computation of a complete Toeplitz matrix with the original random number sequence into multiple sub-operations, effectively reducing the consumption of logical resources and enabling real-time generation of quantum random numbers.Table 2 summarizes the logical resource usage of each work section.
\begin{table}[ht]
\caption{LUTs Usage Quantity. The table shows the LUTs usage ratio of main modules.}
\begin{center}
\setlength\arrayrulewidth{1pt}
\begin{tabular}{c c  } 
 \hline
Module & LUTs Usage \\ [0.5ex] 
 \hline
 ADC & 1083 \\

DDR3 & 12405\\

PCIe & 15814 \\

Toeplitz Extractors & 95736\\

 Others &28481 \\\hline
\end{tabular}
\end{center}
\vspace{-1.5em}%
\end{table}

Each clock cycle we calculate the multiplication between a sub-matric and a subsequence, as shown in equation (5),The m×n toeplitz matrix is divided into n/k (n is an integer multiple of k) sub-matrices having a size of m×k, so that the entire toeplitz matrix processing is split into n/k steps. Each step is a multiplication calculation of m×k sub-matrices and k-bits raw random number sequence, which occupies one clock cycle.
\[\left[ {\begin{array}{*{20}{c}}
{\mathop s\nolimits_m }&{\mathop s\nolimits_{m + 1} }& \cdots &{\mathop s\nolimits_{m + n - 2} }&{\mathop s\nolimits_{m + n - 1} }\\
{\mathop s\nolimits_{m - 1} }&{\mathop s\nolimits_m }& \cdots &{\mathop s\nolimits_{m + n - 3} }&{\mathop s\nolimits_{m + n - 2} }\\
 \vdots & \vdots & \ddots & \vdots & \vdots \\
{\mathop s\nolimits_2 }&{\mathop s\nolimits_3 }& \cdots &{\mathop s\nolimits_n }&{\mathop s\nolimits_{n + 1} }\\
{\mathop s\nolimits_1 }&{\mathop s\nolimits_2 }& \cdots &{\mathop s\nolimits_{n - 1} }&{\mathop s\nolimits_n }
\end{array}} \right] \times \left[ {\begin{array}{*{20}{c}}
{\mathop d\nolimits_1 }\\
{\mathop d\nolimits_2 }\\
 \vdots \\
{\mathop d\nolimits_{n - 1} }\\
{\mathop d\nolimits_n }
\end{array}} \right] = \]
\[\left[ {\begin{array}{*{20}{c}}
{\mathop s\nolimits_m }&{\mathop s\nolimits_{m + 1} }& \cdots &{\mathop s\nolimits_{m + k - 1} }\\
{\mathop s\nolimits_{m - 1} }&{\mathop s\nolimits_m }& \cdots &{\mathop s\nolimits_{m + k - 2} }\\
 \vdots & \vdots & \ddots & \vdots \\
{\mathop s\nolimits_2 }&{\mathop s\nolimits_3 }& \cdots &{\mathop s\nolimits_{k + 1} }\\
{\mathop s\nolimits_1 }&{\mathop s\nolimits_2 }& \cdots &{\mathop s\nolimits_k }
\end{array}} \right] \times \left[ {\begin{array}{*{20}{c}}
{\mathop d\nolimits_1 }\\
{\mathop d\nolimits_2 }\\
 \vdots \\
{\mathop d\nolimits_{k - 1} }\\
{\mathop d\nolimits_k }
\end{array}} \right] +  \cdots  + \]

\begin{equation}
    \left[ {\begin{array}{*{20}{c}}
{\mathop s\nolimits_{m + n - k} }&{\mathop s\nolimits_{m + n - k + 1} }& \cdots &{\mathop s\nolimits_{m + n - 1} }\\
{\mathop s\nolimits_{m + n - k - 1} }&{\mathop s\nolimits_{m + n - k} }& \cdots &{\mathop s\nolimits_{m + n - 2} }\\
 \vdots & \vdots & \ddots & \vdots \\
{\mathop s\nolimits_{n - k + 2} }&{\mathop s\nolimits_{n - k + 3} }& \cdots &{\mathop s\nolimits_{n - 1} }\\
{\mathop s\nolimits_{n - k + 1} }&{\mathop s\nolimits_{n - k + 2} }& \cdots &{\mathop s\nolimits_n }
\end{array}} \right] \times \left[ {\begin{array}{*{20}{c}}
{\mathop d\nolimits_1 }\\
{\mathop d\nolimits_2 }\\
 \vdots \\
{\mathop d\nolimits_{n - 1} }\\
{\mathop d\nolimits_n }
\end{array}} \right] = \left[ {\begin{array}{*{20}{c}}
{\mathop a\nolimits_1 }\\
{\mathop a\nolimits_2 }\\
 \vdots \\
{\mathop a\nolimits_{m - 1} }\\
{\mathop a\nolimits_m }
\end{array}} \right]
\end{equation}

The ISE (Integrated Software Environment) is applied to complete the above design in the FPGA of model 7k325t-fbg676. Operations in FPGAs are typically implemented through look-up tables (LUTs). Although the internal programmable logic resources of FPGAs have been greatly increased in recent years, for large-scale matrix operations, the shortage of FPGA logic resources is still the biggest problem encountered in real-time implementation of Toeplitz-hashing extractor, which is also the key problem need to be solved in this work. Logical resource occupation with different numbers of channels is compared based on ISE. It can be found that the LUTs occupied by PCI-E and functional modules in fact unchanged as the number of all FPGA LUTs as the number of parallel channels increases, approximately 28.4$\%$ of all FPGA LUTs.For four-way parallel post-processing with scale of about 1700*2500, this setup makes reasonable use of resources and the Toeplitz matrix post-processing algorithm for each channel occupies approximately 12$\%$.Figure 1 shows the change in hardware resources as they are occupied by multiple post-processing operations.

\begin{figure}[htb!]
\centering\includegraphics[width=7cm]{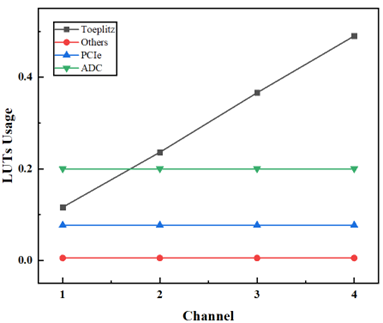}
\caption{Post-processing logic resource utilization}
\end{figure}

\section{Parallel post-processing of QRNG}

The experimental scheme is shown in Fig.2. A single-mode laser operating at a center wavelength of 1550 nm acts as the local oscillator. Single-mode continuous-wave beam with a power of 6.4 mW enters into one port of a balanced beamsplitter, while the other input port was blocked to ensure only the vacuum state coupled in. The vacuum field and the LO interfered in the second 50/50 beamsplitter and was output with balanced power. The outputs are detected simultaneously by BHD (PDB480C, Thorlabs Inc., Newton, MA, USA) to cancel the excess noise of the LO as well as amplify the quadrature fluctuations of the vacuum state.

\begin{figure}[ht!]
\centering\includegraphics[width=12cm]{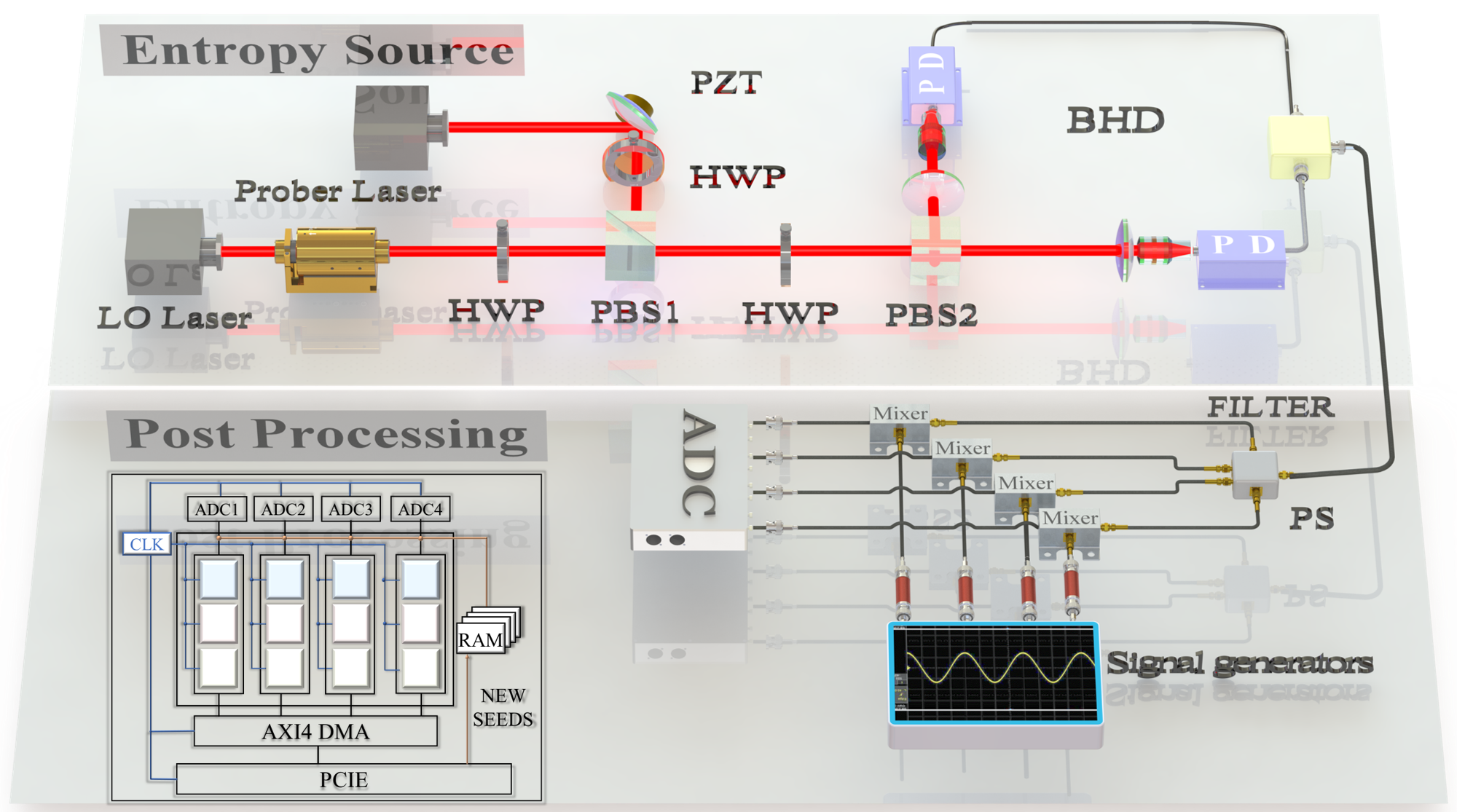}
\caption{Experimental implementation of real-time parallel post-processing of vacuum state-based QRNG. The entropy source part uses a vacuum-based scheme to generate a randomly fluctuating voltage signal. The post-processing part uses mixing, filtering, analog-to-digital conversion, and Toeplitz post-processing to obtain random bits. HWP, half wavelength plate; PBS, polarization beam splitter; BHD, balanced homodyne detector; SG, signal generator; ADC, analog to digital converters.}
\vspace{-1.5em}%
\end{figure}

Fig. 3 is a frequency spectrum diagram of the output voltage of the homodyne photodetector.Based on discussion in last section, four sub-entropy sources with bandwidth are exploited to generate quantum random numbers. As shown in Fig.2, quantum sideband modes centering at distinct analysis frequencies (200MHz, 400MHz, 600MHz and 800 MHz) are extracted from four portions of the homodyne photoelectric signalThe SNR of the chosen four bands is greater than 10 dB to ensure that quantum noise fluctuations account for a large proportion of the measurement results relative to the effects of classical noise. Mixers and filters are used to achieve frequency band extraction. Reference signals for mixing are provided by a signal generator (SG). Four analog to digital converters (ADCs) with sampling frequency of 250 MHz and a sampling accuracy of 16 bits are employed to sample the signals under four frequency-mode. According to Nyquist’s theorem and the sampling rate of ADC, the bandwidths of the four sub-entropy source are 98 MHz, which is the cut-off bandwidth of the low-pass filter(BLP-100+).Raw data sourced from the four modes are parallelly post-processed in an FPGA (xc7k325t-2ffg676) in real-time,and multi-channel data are aggregated and input to a computer through the PCI-E interface.
\begin{figure}[ht!]
\centering\includegraphics[width=8cm]{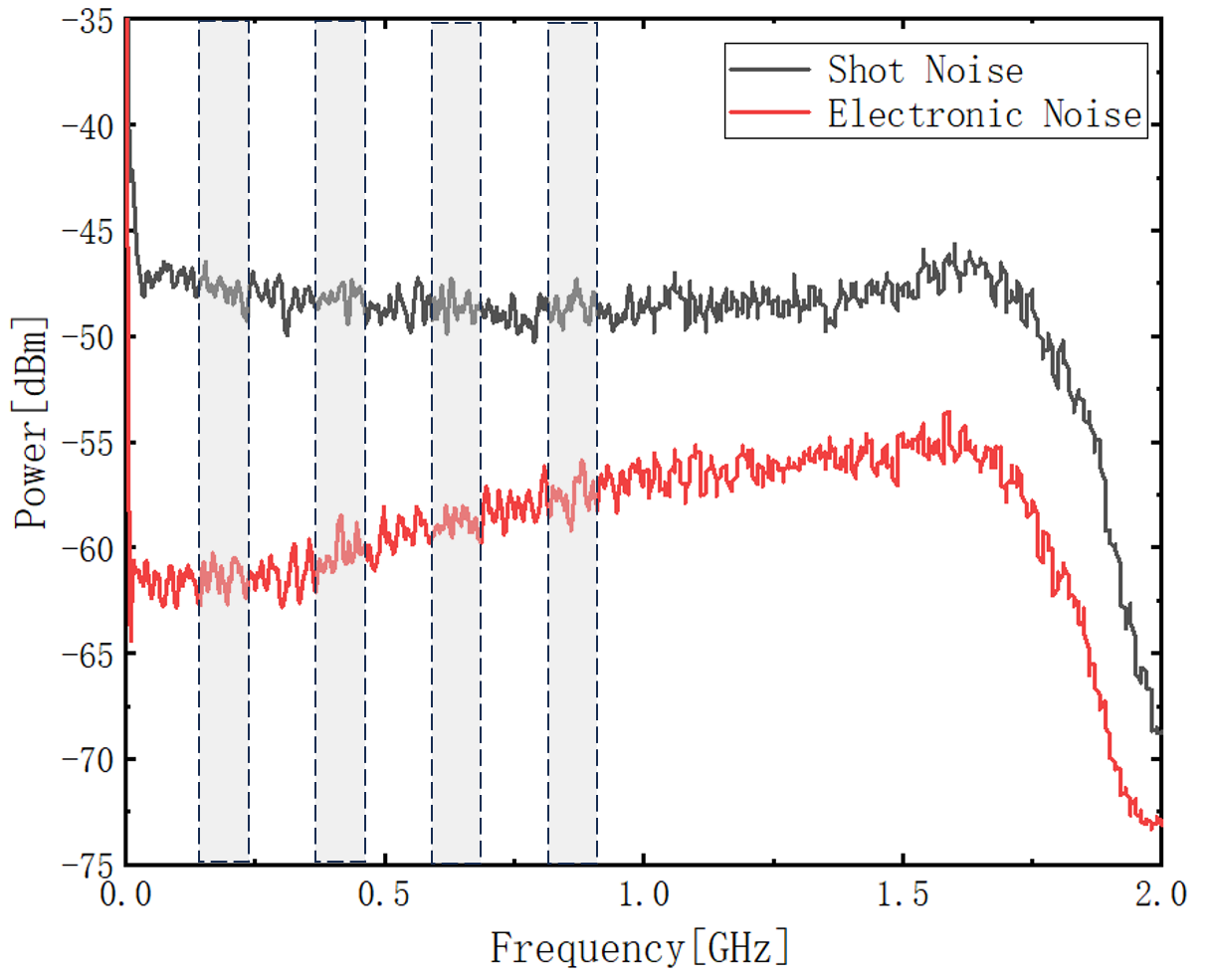}
\caption{The red line is the power spectrum of the electronic noise, the black line is the power spectrum of the shot noise.}
\vspace{-1.5em}%
\end{figure}

\section{Post-processing with real-time seed update and system data flow}

\subsection{\textbf{Sub-seed Generation and system data flow}}

In practical applications, reusing the same seed for post-processing requires that the original data be mutually independent, a condition generally not satisfied by general input raw data. In our work, we address this challenge by randomly selecting one seed from multiple seeds for each post-processing task, thereby achieving a relative independence between successive operations. Additionally, to maintain high security standards in post-processing, we opt for real-time updates of seeds sourced externally to prevent excessive escalation of security parameters.
Since the matrix sizes of the four Toeplitz-hashing extractors are different, in the following description of the specific algorithm, m, n and k are used to illustrate the general case. In one clock cycle, the FPGA completes the construction of a Toeplitz submatrix and the processing of k-bit raw random numbers. After n/k clocks, all n/k sub-matrices are used. The calculated m-bit vector is the result of post-processing of the entire Toeplitz matrix. During n/k clocks, the ADC acquires n-bits raw random numbers, and the FPGA processes the n-bit raw random number in real-time and output m-bit true random numbers.
Before delving into the specific scheme for real-time updates of external seeds, it is essential to address the generation of sub-seeds. Sub-seeds are derived from seeds and are utilized in each post-processing task with a length of m+k-1 bits. Traditional methods for generating sub-seeds involve storing seeds in registers and sequentially shifting strings to produce them. Registers shift rightward by k bits each time, selecting the lowest m+k-1 bits. Each seed of length m+n-1 can generate n/k sub-seeds. However, our approach differs from traditional methods due to security considerations. Considering the large scale of post-processing matrices and the necessity for random selection of multiple seeds per channel, adopting the conventional approach for sub-seed generation would consume significant registers and logic resources, leading to routing congestion and timing hazards.
Here, we propose a novel method. We outline the flow of data within the system and detail the generation of sub-seeds. As illustrated in Figure 4, we have constructed the overall system of a quantum random number generator post- processing module, utilizing PCIe as the transmission interface. The PCIe module serves as the master, communicating with other slave modules via the AXI4 bus. An AXI4 arbiter determines the slave object to interact with the PCIe module during each transaction. When the security parameters of post-processing reach a predefined threshold, new true random numbers are written into the first-level memory for generating the original matrix. The second-level memory is utilized to store the seeds for constructing sub-matrices, with sub-seeds generated from the original seeds. A memory controller manages the readout of the first-level memory and the writing control of the second-level memory. Each of the four independent Toeplitz modules is equipped with corresponding first and second-level memories to provide seeds for real-time updates. True random numbers generated by the Toeplitz module are written into DDR3 memory and transmits them to the host computer.

\begin{figure}[ht!]
\centering\includegraphics[width=16cm]{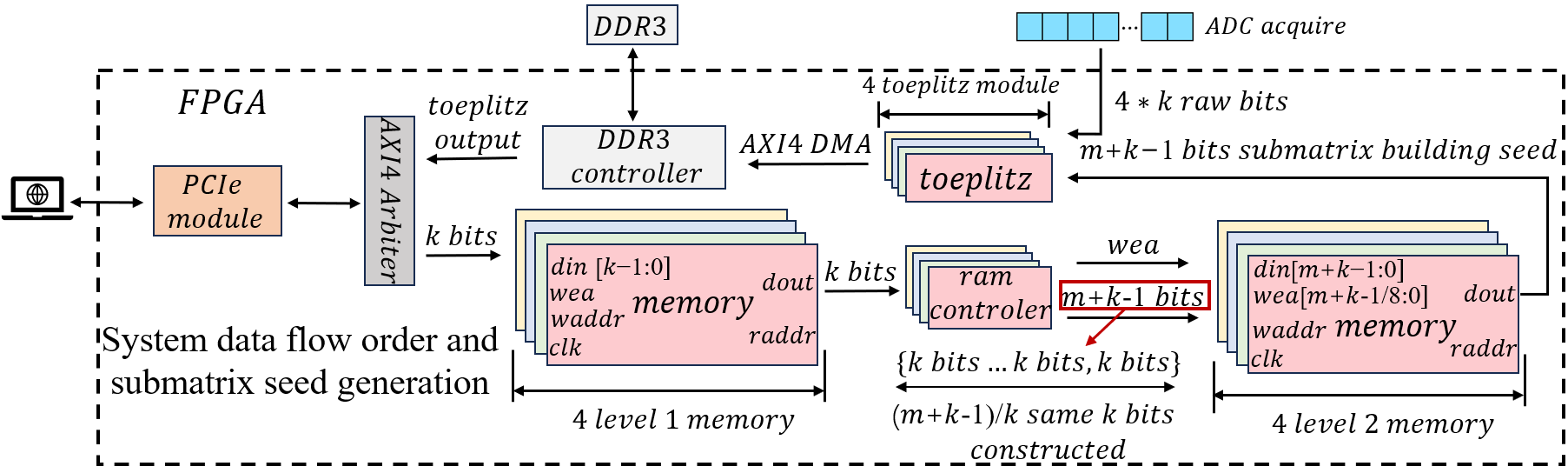}
\caption{The system architecture for data flow and submatrix seed generation is organized into five key modules: PCIe, AXI4 bus arbitration, DDR3 storage, Toeplitz post-processing, and submatrix seed generation. The PCIe module, comprised of the PCIe PHY and controller configured for AXI4 interface, enables data exchanges between the host and peripheral devices. Users can either retrieve random numbers from DDR3 over PCIe to the host or update the system with verified random seeds into the initial four memory units. The DDR3 module integrates external DDR3 memory with an AXI4-configured controller for efficient data management. The submatrix seed generation is achieved through four sets of memory units and their controllers, linked to four parallel Toeplitz modules, which direct the generated random numbers into DDR3 via DMA for optimal .}
\vspace{-1.5em}%
\end{figure}

We propose an innovative approach that utilizes FPGA memory resources to address these challenges. Instead of using shift logic operations of registers, we employ read and write operations between two memory modules. The first-level memory has a width of k bits and a depth of b×(m+n-1)/k, storing b seeds. The second-level memory, with a width of m+k-1 bits and a depth of b×n/k, uses byte-enable (8 bits) for write enable. As shown in Figure 5 ,this architecture abstracts memory into one-dimensional data blocks, with each block in the first-level memory having a length of k and each block in the second-level memory having a length of m+k-1.To generate sub-seeds, we sequentially read k bits random numbers from the first-level memory based on different address ranges and write them to corresponding addresses in the second-level memory, dynamically adjusting the write enable. Once all data at the same address in the second-level memory is filled, a sub-seed is generated. This process repeats until n/k sub-seeds are created. Finally, repeat the entire process b times to complete the generation of sub-seeds for b seeds.The sub-seed generation consumes logic resources only during write enable shifting and address changes.The Toeplitz module retrieves corresponding seeds from the second-level memory for post-processing, and the generated random numbers are written into the external DDR3 module. This streamlined approach optimizes resource utilization while maintaining robustness and efficiency in random number generation on FPGA platforms.

\begin{figure}[ht!]
\centering\includegraphics[width=16cm]{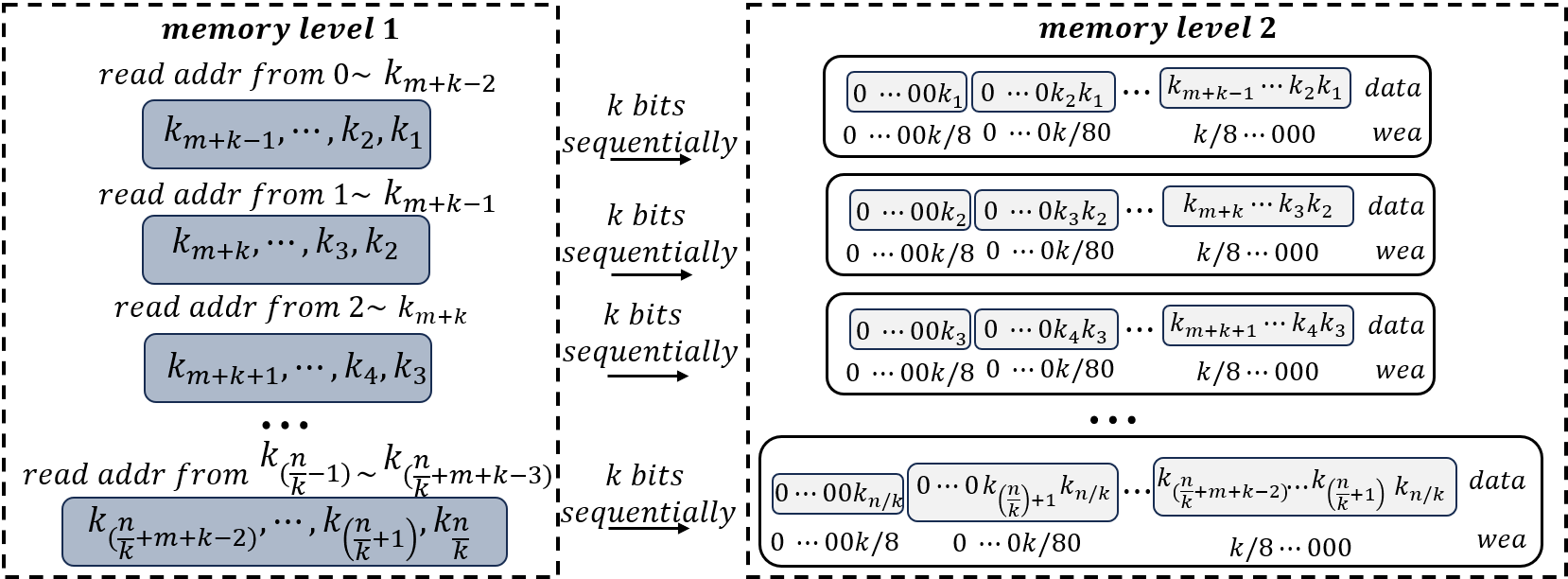}
\caption{The specific method for sub-seed generation involves the controller sequentially reading the updated k-bit random seed from the first-level memory and controlling the write enable of the second-level memory to fill the corresponding locations with the k-bit random data. This process is repeated until the construction of the m+k-1 bit sub-seed is completed.}
\vspace{-1.5em}%
\end{figure}

\subsection{\textbf{Post-processing with Randomizable Seeds}}

As mentioned earlier, we store b sets of sub-seeds in the secondary memory to generate subsets forming matrices. Each set comprises n/k sub-seeds,and each set of sub-seeds begins at the memory's initial address,y=0,1,2$\cdots$,b-2,b-1.As shown in Figure 6, The constructed LSFR (Linear Feedback Shift Register) circuit generates $x$-bit random numbers at each clock cycle, representing values in the range of 0 to $2^{x}$-1. These values are chosen such that $2^{x}$ is a multiple of b, thereby dividing the representable data range into b equal parts. Because the values of the random numbers are evenly distributed across the regions, after completing one hashing of an m×n matrix, we perform a numerical judgment on the random numbers generated by LSFR. We determine the interval in which the numerical value falls and, based on that interval, select a different random seed's starting address in the memory for the new toeplitz post-processing.

 As depicted in Figure 6(a), we implement the toeplitz post-processing through three parallel pipeline modules. In the first stage, seeds for constructing submatrices are obtained from memory either through random selection (after completing one post-processing iteration) or through sequential selection. The second stage is responsible for the construction of submatrices and their subsequent multiplication. When constructing a submatrix, from the j bit to the j+m-1 bit of the selected seeds fill the jth column of the submatrix. The value of j ranges from 1 to k.After the construction of the submatrix, it undergoes operations with the input k-bit original random numbers, wherein the process can be decomposed into multiplication and addition. For binary matrices, we employ the bitwise AND operation and bitwise XOR operation to execute the multiplication and addition processes of the matrix. This constitutes a one-cycle operation that adheres to the timing requirements. The corresponding logic circuit implementation is illustrated in Figure 6(b). The third stage is register accumulating module. In the first cycle, we store the result of a single submatrix processing in registers, and in each subsequent cycle, we accumulate the value of the register with the result of the single submatrix processing to get the final result in the  n⁄k cycle. The corresponding logic circuit implementation is illustrated in Figure 6(c).

 \begin{figure}[ht!]
\centering\includegraphics[width=16cm]{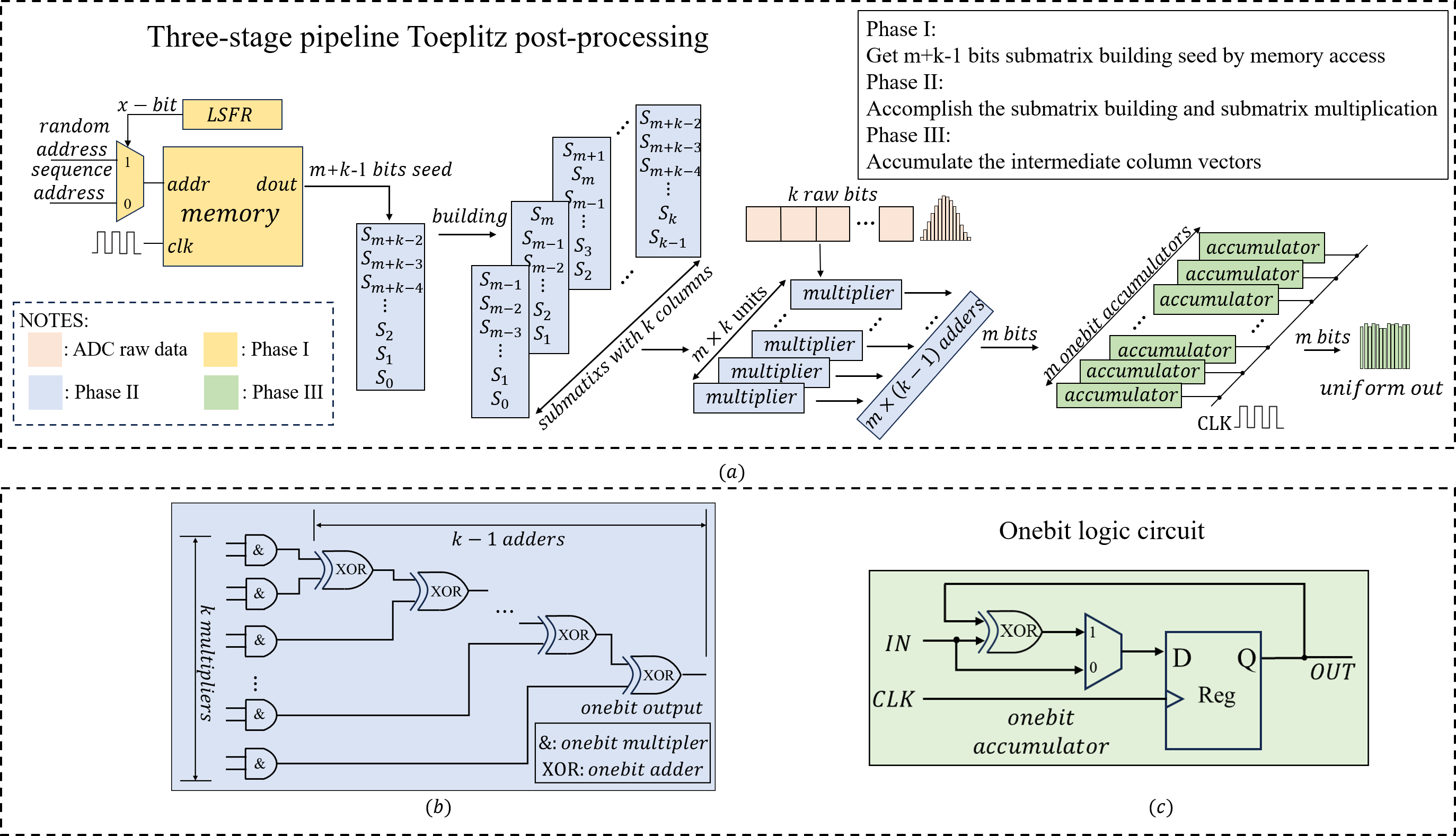}
\caption{Illustrates the comprehensive schematic for the three-tier parallel Toeplitz post-processing architecture. This process encompasses: 1) Memory access to fetch sub-seeds, leveraging LFSR-generated randomness for determining a random start address for sub-seed retrieval, followed by sequential address incrementation to continuously access sub-seeds from the same set. 2) Construction of sub-matrices and their multiplication with the input original random sequence, following Toeplitz conventions to facilitate single-cycle multiplication operations. 3) Cumulative addition of the sub-matrix multiplication outcomes. Figure 6(b) details the logic circuitry for the sub-matrix multiplication, whereas Figure 6(c) depicts the logic circuitry for the cumulative addition of the multiplication results from sub-matrices.}
\vspace{-1.5em}%
\end{figure}

\subsection{\textbf{}\textbf{OPTIMIZATION OF TIMING IN FPGA}
}

In this work, we used a variety of methods to optimize the timing of the FPGA to achieve the optimal processing results. (1) The ADC acquisition clock of the four channels is C, and the ADC sampling accuracy of each channel is a. For FPGA, the high sampling frequency of the ADC will bring greater timing pressure, so we use FIFO to convert the original random sequence with a sampling rate of C and a sampling accuracy of a into a random sequence with clock frequency J and k, k=Ca⁄J. (2) By default, the input signal of each FPGA is connected to the input buffer as a driver. On top of this, we added a global buffer. Their input can be a clock generated by a phase-locked loop, the output of other buffers, general internal signal lines, etc. The output of the global buffer can be connected to each clock input point on the chip. Xilinx 7 Series FPGA have 32 global buffers. We plug the clock and certain signal lines that are too high to fan out into the global buffer to improve their drive capability and reduce fan out. (3) Since the entire Toeplitz post-processing module runs in a pipeline mode, the fan-out of registers will be large, so we use the method of register copying to reduce the fan-out of registers and optimize the timing of post-processing. The following table represents the optimized timing results. It can be seen that the timing analysis of the four-channel Toeplitz post-processing alone shows that our design is timing convergent. With the addition of the remaining blocks, more cross-clock domain paths and more complex control signals were introduced, timing closure was difficult, and the critical path for constraint failure was to convert the control signal of the ADC trigger mode, which did not affect the data acquisition and the entire post-processing process.

\begin{table}[ht]
\caption{Timing optimization results.}
\begin{center}
\setlength\arrayrulewidth{1pt}
\begin{tabular}{c c c c} 
 \hline
 module & LUTs Usage& Timing Score& WorstCaseSlack \\ [0.5ex] 
 \hline
 Toeplitz Only & 48$\%$ & 0 & 0.623ns  \\
 With Others  & 76.4$\%$ & 1487 & -0.066ns  \\\hline
\end{tabular}
\end{center}
\vspace{-1.5em}%
\end{table}

\section{TESTING AND ANALYSIS}
To ensure the independence between the channels,the correlation coefficient and mutual information were calculated for the quantum random numbers obtained from two pairs of channels in a real-time and efficient post-processing system. As shown in Figure 7,the correlation coefficient and mutual information can be used to characterize the correlation between the selected sets of random bit sequences. The correlation coefficients and mutual information between each pair of channels are all below the magnitude of $10^{-3}$and $10^{-6}$respectively, indicating an extremely low level of correlation between different post-processing channels. This shows that random bits can be generated with high quality in each frequency band, and the correlation between the four channels can be ignored. To verify the randomness of the random number streams extracted from each subentropy source, as well as their independence, we constructed 64×64 random binary bitmap images using four binary sequences. As shown in Figure 8, the individual bitmap images from all four data channels did not exhibit any noticeable patterns or biases.
\begin{figure}[ht!]
\centering\includegraphics[width=12cm]{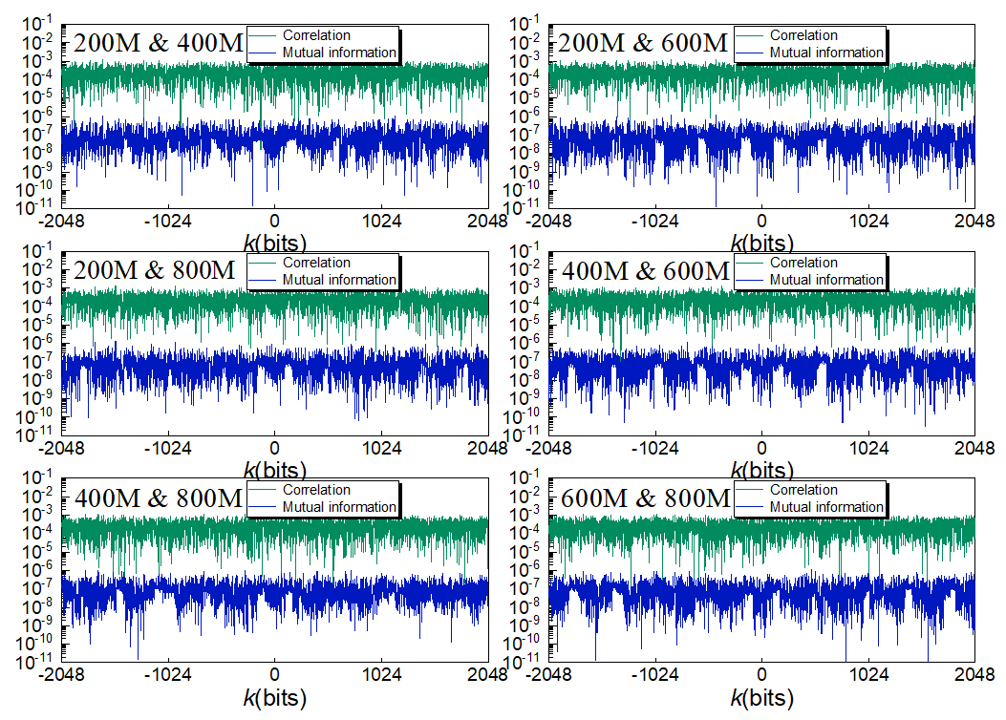}
\caption{Channel's cross-correlation and mutual information .The mutual information and cross-correlation among samples are at a very low magnitude, demonstrating a high level of independence.}
\end{figure}
In order to verify the randomness of the random bits strictly, we firstly collected the output sequences of four channels separately and performed 15 statistical tests using the standard NIST Statistical Test Suite[18]. The significance level is set as $\alpha$ = 0.01 and 1000 sequences with 1M bits are subjected to the test. Then cumulated sequences from mixing of the four outputs from the four paths are tested under the same conditions.  As shown in Fig. 9, all P values are greater than 0.01, and the minimum pass rate for each statistical test is also within the confidence interval of 0.981 to 0.995.Additionally, we used the TestU01 and DieHard test suites for testing, with the results shown in Figures 10 and 11. The output of the extractor successfully passed all standard statistical tests.These tests identify the reliability of each Toeplitz-hashing extractor for every quantum sideband mode and demonstrate the randomness of the ultimate random numbers generated by the parallel QRNG.
\begin{figure}[ht!]
\centering\includegraphics[width=14cm]{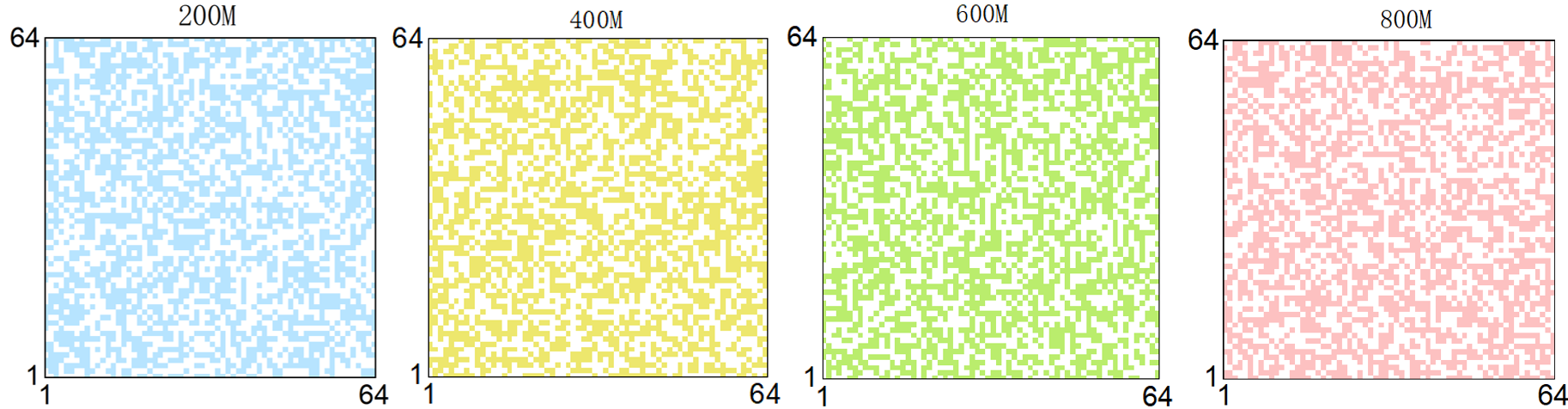}
\caption{The individual bitmap images from all four data channels.}
\end{figure}

\begin{figure}[ht!]
\centering\includegraphics[width=10cm]{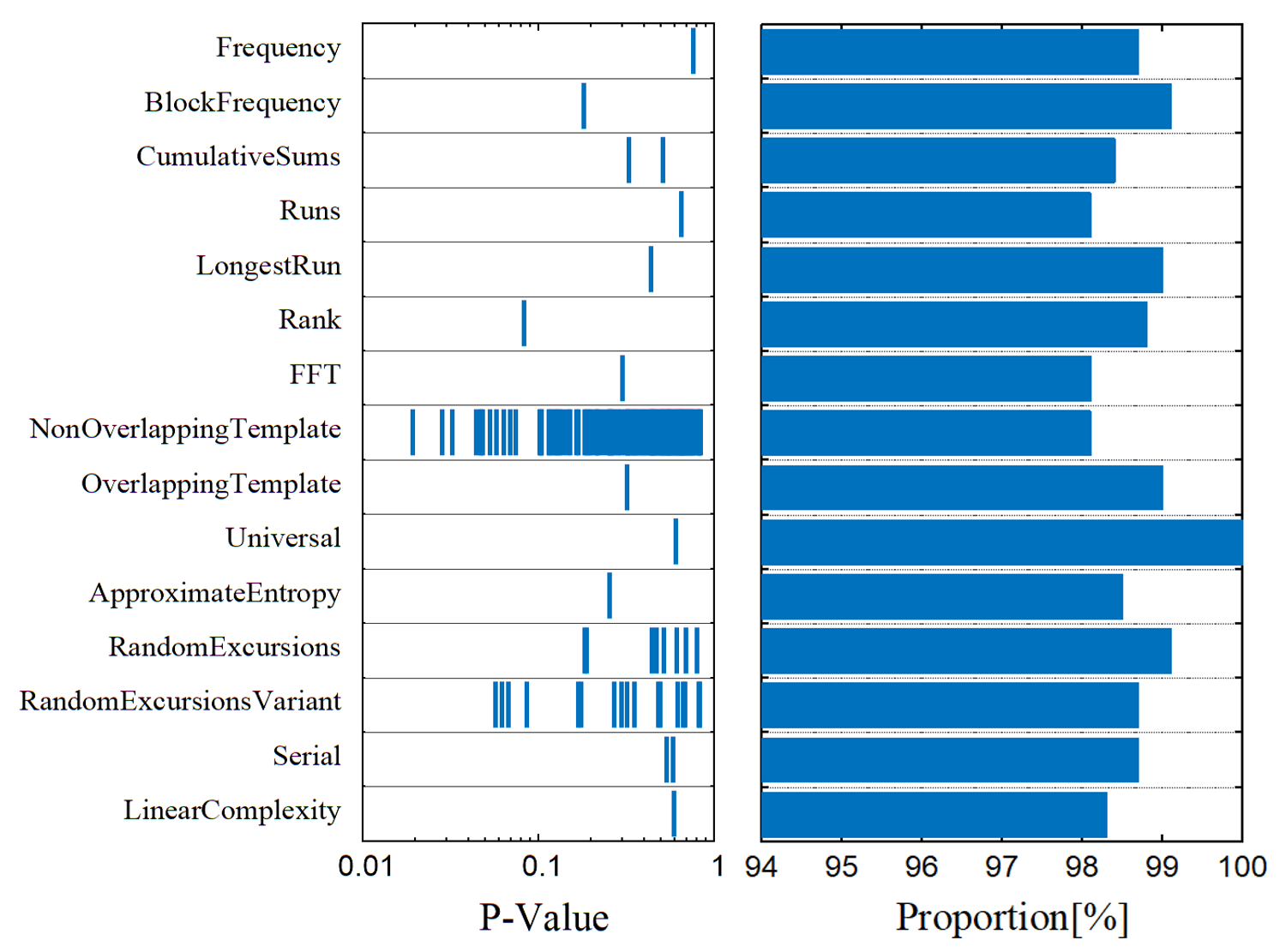}
\caption{Typical results of standard NIST Statistical Test Suite. .}
\end{figure}

\begin{figure}[ht!]
\centering\includegraphics[width=10cm]{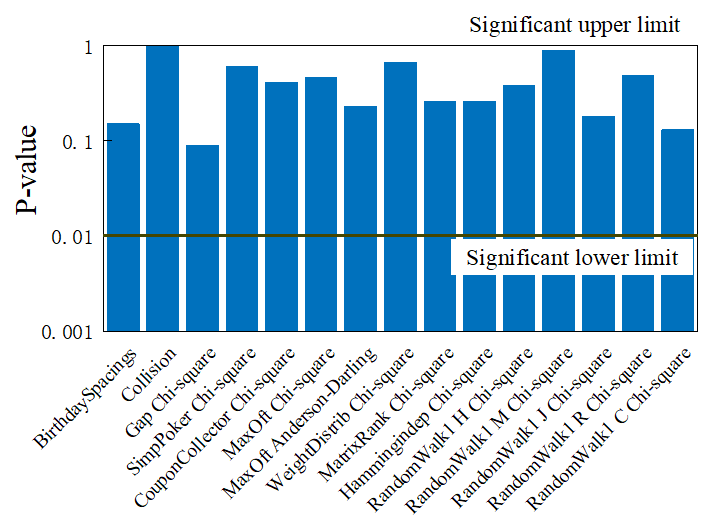}
\caption{Typical results of standard DieHard Test Suite. }
\end{figure}

\begin{figure}[ht!]
\centering\includegraphics[width=10cm]{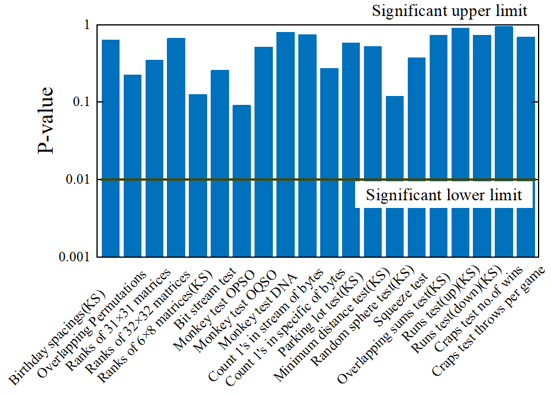}
\caption{Typical results of standard DieHard Test Suite .}
\end{figure}

\section{Conclusion}

In this work, we make efforts to coordinate the logic resources inside a FPGA to realize real-time post-processing as high-speed as possible. Based on progressive elaboration with regards to security, parallel computing of four toeplitz matrices with a dimension about 1700*2500 can make full use of logical resources the most in the environment of xilinx Kintex7-325t. Under these operating conditions, we used fewer resources to build a 4-channel post-processing block, and the toeplitz matrix size was 3 times larger than in previous work.To simultaneously implement toeplitz post-processing of the raw data from the four quantum frequency modes in real-time and high-speed, we elaborately coordinate the logic resource and clock signal and establish a two-layer parallel pipeline construction. In this way, four toeplitz-hashing extractors is realized within one FPGA based on fully and efficiently utilization of the FPGA resource.

 Additionally, this study analyzes and discusses the issue of seed updates, incorporating a process of random seed selection during post-processing to ensure each session is i
 ndependent of the others. This satisfies the prerequisite for reusing seeds. Furthermore, a threshold for security parameters is set, allowing for the rapid update of new random seeds through the PCIe interface when the security parameter reaches the predetermined threshold. Ultimately, on the same hardware platform used in previous work, we achieved a random number generation rate of 11.3Gbps. Not only is this rate higher, but the security is also improved. Moreover, with the use of higher-performance ADCs and post-processing platforms, a random number generation rate exceeding 20Gbps was achieved.

\printbibliography
\cite{1,2,3,PhysRevA.87.062327,Qi:10,Li:19,1950Various,DEEPTHI2009229,article,Trevisan2001ExtractorsAP,Xu_2012,10.1063/1.4922417,10.1063/1.5078547,Guo:19,zkd1111,5961850,e20110819,8966,Drahi2019CertifiedQR,Gehring2021HomodynebasedQR,li2024improved,Bruynsteen_2023,article1,2020QuIP...19..405L,10.1145/73007.73012}

\end{document}